\documentclass[aps,prl,reprint,groupedaddress,floatfix]{revtex4-1}

\usepackage{color}
\usepackage{graphicx}
\usepackage{dcolumn}
\usepackage{bm}
\usepackage{verbatim}
\usepackage{upgreek}
\usepackage{comment}
\usepackage{hyperref}

\begin{document}

\title{Antiferromagnetism in NiO Observed by Transmission Electron
  Diffraction}

\author{J.C. Loudon}
\email{j.c.loudon@gmail.com}

\affiliation{Department of Materials Science and
  Metallurgy, University of Cambridge, Pembroke Street, Cambridge CB2
  3QZ, United Kingdom}

\date{\today}

\begin{abstract}

Neutron diffraction has been used to investigate antiferromagnetism
since 1949. Here we show that antiferromagnetic reflections can also
be seen in transmission electron diffraction patterns from NiO. The
diffraction patterns taken here came from regions as small as 10.5~nm
and such patterns could be used to form an image of the
antiferromagnetic structure with a nanometre resolution.
\end{abstract}


\maketitle

Antiferromagnetic materials have opposed magnetic moments on adjacent
atoms so produce no external magnetic field. They were identified in
1932~\cite{Neel32} and today their main application is for computer
hard drive readers which use the exchange-bias effect~\cite{Dieney91,
  Ijiri98}. The arrangement of the atomic magnetic moments can be
deduced by recording diffraction patterns using radiation sensitive to
the magnetic flux density between the atoms. Neutrons have a magnetic
moment ${\bm \upmu}_n$ (of magnitude $9.65\times10^{-27}$~Am$^2$) and
so feel a force ${\bf F}_n=\bm{\upmu}_n. \nabla {\bf B}$ on passing
through a flux density ${\bf B}$. Neutron diffraction was
first used to detect antiferromagnetism in 1949~\cite{Shull49} and has
since been used extensively to study the structure of
antiferromagnets~\cite{Harrison06}.

Electrons should also be sensitive to antiferromagnetism as they are
charged particles (with charge $-e=-1.602\times10^{-19}$~C) and so
experience a Lorentz force ${\bf F}_e=-e{\bf v} \times {\bf B}$ on
passing through a B-field with a velocity ${\bf v}$. The effect of the
Lorentz force is used extensively in electron microscopy to map
B-fields in ferromagnets~\cite{Midgley01} but our literature survey
found no evidence of its use to examine antiferromagnets.

Antiferromagnetic domains have been imaged before using transmission
electron microscopy but these images did not use magnetic scattering
but relied on the fact that the domains observed were also structural
twins~\cite{Remaut65}. In \textit{Electron Microscopy of Thin
  Crystals}~\cite{Hirsch65} it states `it is not clear whether the
periodicity in the spins [of an antiferromagnet] can give rise to
observable diffraction effects [in a transmission electron diffraction
  pattern]'. Here we show that antiferromagnetic reflections can be
observed in electron diffraction patterns and are about $10^{4}$ times
less intense than the structural Bragg peaks unlike neutron
diffraction where both are of similar intensity~\cite{Harrison06}.

The Lorentz force is the dominant force felt by electrons due to the
magnetic flux density in the specimen at the energies used for
transmission electron microscopy (300~keV here) as shown in
Supp. Info. 1. Unlike neutrons, the magnetic force due to the
electron's dipole moment never dominates. At low energy, the exchange
interaction dominates~\cite{DeWames67} and low energy (32~eV) electron
diffraction patterns have been acquired from the first atomic layer of
an antiferromagnet using this effect~\cite{Menon11}.

The advantage of transmission electron microscopy is its potential to
examine features of the antiferromagnetic structure such as domain
walls at high resolution. The diffraction patterns taken here came
from regions as small as 10.5~nm and a 1~nm resolution should be
possible. For comparison, the resolution of neutron
imaging~\cite{Schlenker94} is 60~$ \upmu$m; low energy electron
diffraction, 10~nm~\cite{Menon11} and photoemission electron
microscopy, 20~nm~\cite{Arai12}.

In this experiment, we acquired electron diffraction patterns from
single crystal NiO. It was one of the first antiferromagnets
investigated by neutron diffraction so its magnetic structure is well
characterised~\cite{Roth58}. Its crystal structure is based on the
face-centred cubic, sodium chloride structure~\cite{Slack60} with a
lattice parameter of $a=4.18$~\AA. Antiferromagnetic order occurs
below the N\'{e}el temperature, $T_N=523$~K and consists of the
ferromagnetic alignment of the magnetic moments of the Ni ions in one
set of (111)-type planes with the moments in alternate (111) planes
being antiparallel~\cite{Roth58} (see Supp. Info. 2). Experimental
measurements of the magnetic moment give values between 1.77 and
2.2$\pm$0.2$\mu_B$ per Ni ion~\cite{Kwon00} and the usual picture is
that each Ni$^{2+}$ ion has its spin-only magnetic moment of $2\mu_B$
although this is questioned in ref.~\cite{Kwon00}.

Antiferromagnetic order is accompanied by a slight rhombohedral
distortion~\cite{Slack60} which compresses the lattice along the [111]
axis normal the ferromagnetic (111) planes. This distortion does not
lead to extra reflections and we index diffraction patterns using the
conventional cubic coordinates. In Supp. Info. 3, it is shown that if
positions in reciprocal space of the structural Bragg reflections are
denoted ${\bf G}=(hkl)$ and the antiferromagnetic modulation in the
local magnetisation is written ${\bf M}={\bf M}_0\cos(2\pi {\bf q.r})$
(where ${\bf r}$ is a position vector, ${\bf q}$ is the wavevector of
the modulation and ${\bf M}_0$ is the sublattice magnetisation),
antiferromagnetic reflections occur at positions ${\bf Q}={\bf
  G}\pm{\bf q}$ in diffraction patterns acquired using radiation
sensitive to magnetism. For NiO, $h$,$k$ and $l$ must all be odd or
all even numbers. The antiferromagnetic wavevector points in one of
four possible $\{111\}$ directions and each will generate different
antiferromagnetic reflections. For example, the lowest order
reflections generated by ${\bf q}={1 \over 2}(111)$ will be ${\bf
  Q}=\pm{1 \over 2}(111)$, $\pm{1 \over 2}(11\overline{3})$, $\pm{1
  \over 2}(1\overline{3}1)$, $\pm{1 \over 2}(\overline{3}11)$ but not
${1 \over 2}(113)$ as there is no allowed structural reflection, ${\bf
  G}$, from which this could originate.

In the absence of an applied magnetic field, the magnetic moments of
the Ni ions, $\bm {\upmu}$, point in one of the three
$\langle11\overline{2}\rangle$ directions (called `easy axes') which
lie in the ferromagnetically aligned (111)
planes~\cite{Baruchel81}. When the direction of the antiferromagnetic
order changes, there are two types of domain boundary: a twin (T-type)
boundary occurs when the magnetic order changes to a different set of
(111) planes and the accompanying distortion generates a
crystallographic twin. The spin (S-type) domain boundary occurs where
the same set of (111) planes remains ferromagnetic but the magnetic
moments point in a different direction (see Supp. Info. 2).

For most of this experiment, the microscope was used in its normal
operating mode where the objective lens applies a 2.8~T magnetic field
to the sample parallel to the electron beam. Above 1.54~T, a
`spin-flop' takes place in single crystal NiO where the spins are
realigned so they remain in the same zone as ${\bf q}$ but point
normal to the applied field~\cite{Saito80}. For the [112]-type zone
axes used here, the direction of the magnetic moments in the
spin-flopped configuration will be of the $[1\overline{1}0]$-type. In
addition, the spins will be canted in the direction of the field by an
angle of $9^\circ$ (calculated from the susceptibility measurements in
ref.~\cite{Singer56}). The induced spin flop is an advantage for the
purpose of seeing antiferromagnetic ${\bf Q}={1 \over
  2}(11\overline{1})$-type reflections, as they have maximum intensity
when the magnetic moments are normal to the incoming beam. The spin
flop also avoids the possibility of the magnetic moments being
parallel to the beam which would render the antiferromagnetic
reflections invisible. We have also taken diffraction patterns in a
low-field configuration as discussed later.

In Supp. Info. 3, the phase object approximation is used to derive the
following formula for the intensity of an antiferromagnetic reflection
at position ${\bf Q}$, $I_{\bf Q}$, relative to the 000 beam, $I_0$,
for electron scattering:

\begin{equation}
\label{I}
{I_{{\bf Q}} \over I_0}=\left({e \over 2h}{\mu_0\mu_B \over \Omega}{t \over
    Q}\,(\bm {\widehat{\upmu}}\times {\bf \widehat{Q}}).{\bf \widehat{z}}\,F({\bf Q})\right)^2
\end{equation}

where $h$ is Planck's constant, $\Omega$ is the unit cell volume and
$t$ is the specimen thickness. $\bm{\widehat{\upmu}}$ is a unit vector
in the direction of the magnetic moments, ${\bf\widehat{Q}}$ is a unit
vector in the scattering direction and ${\bf\widehat{z}}$ is a unit
vector in the direction of the incident electron beam. The structure
factor is given by $F({\bf Q})\equiv\sum_jn_jf_j({\bf Q})e^{2\pi i{\bf
    k.Q_j}}$ where the sum is over all the atoms in one unit cell,
$n_j$ is the number of unpaired electrons associated with atom $j$ and
$f_j({\bf Q})$ is the magnetic form factor for atom $j$. The form
factor is the same as for neutron diffraction because the same
electrons generate the flux density. We use the values given in
ref.~\cite{Alperin61}. The calculated intensity ratios are tabulated
in Supp. Info. 4 for a 100~nm thick specimen in the
$\left<112\right>$-type zone axes for $\bm{\upmu}$ in
$\langle1\overline{1}2\rangle$-type easy directions as well as the
spin-flopped $[1\overline{1}0]$-type directions.

The NiO single crystals used in this experiment were supplied by
Pi-Kem Ltd and were of size $5\times 5 \times 0.5$~mm with the largest
surface being (111). They were prepared for electron microscopy by
thinning in the [111] direction by mechanical polishing followed by
argon ion beam thinning using a Gatan Precision Ion Polishing System
(PIPS). Electron microscopy was conducted using a Philips CM300
transmission electron microscope equipped with a field-emission gun
operated at 300~kV.

The advantage of a crystal thinned in the [111] direction is that the
[112], [121] and [211] zone axes are accessible with a double-tilt
holder as each lies at $19.5^{\circ}$ to [111]. This ensures that
antiferromagnetic reflections are seen in at least one of these zones
for any ${\bf q}$. The disadvantage is that the (111) surface is
unstable to surface reconstructions which can also produce extra
reflections in a diffraction pattern~\cite{Ciston10}.

Electron diffraction patterns were acquired using Ditabis imaging
plates which have a high dynamic range of 2 million grey levels. They
were not filtered to remove inelastic scattering as the Gatan Imaging
Filter on this microscope does not allow energy-filtered images to be
acquired on imaging plates. The patterns were acquired with exposures
between 0.2 and 200~s at camera lengths of 740--3900~mm from regions
10.5--880~nm in diameter. The shortest exposure in which
antiferromagnetic reflections could be seen was 0.24~s.

\begin{figure}
\includegraphics[width=60mm]{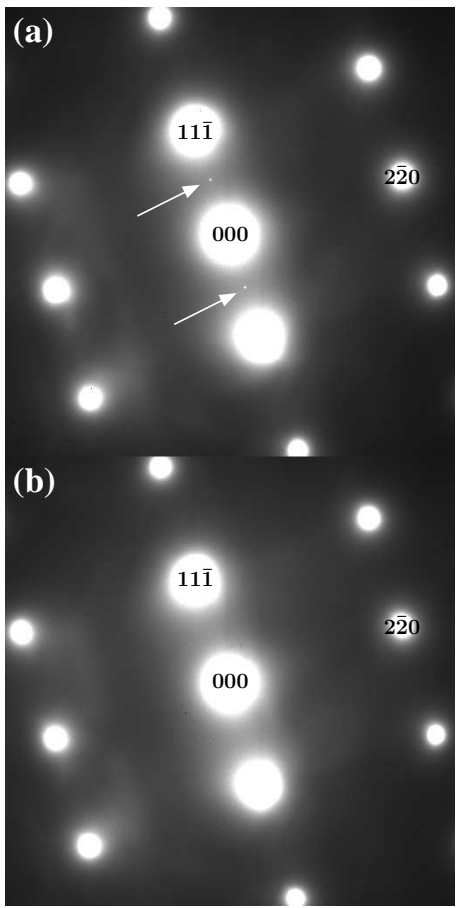}
\caption{\label{fig1} (a) Electron diffraction pattern from the [112]
  zone axis taken at room temperature from a region of diameter 880~nm
  with an exposure of 70~s. $\pm{1 \over 2}(11\overline{1})$
  superlattice are indicated by arrows. (b) Electron diffraction
  pattern from the same region taken at 563~K, above the N\'{e}el
  temperature $T_N=523$~K, showing that the superlattice reflections are
  no longer present.}
\end{figure}

We first used polarised light microscopy to visualise the rhombohedral
(T-type) domains (Supp. Info. 5) showing they were 2--80~$\upmu$m in
size. We then acquired electron diffraction patterns and
Fig.~\ref{fig1}(a) shows a pattern taken at room temperature from the
[112] zone axis with superlattice reflections at ${\bf Q}={1 \over
  2}(11\overline{1})$. The full-width-half-maximum of the superlattice
reflections was the same as the structural reflections to an upper
limit of 0.2\% of $|g_{111}|$ indicating very good long-range
order. No superlattice reflections were seen in the same area from the
[121] and [211] zone axes, consistent with the area being a single
antiferromagnetic domain.

To investigate whether the reflections were generated by
antiferromagnetism, the sample was heated above the N\'{e}el
temperature ($T_N=523$~K) using a Philips PW6592 heating
holder. Fig.~\ref{fig1}(b) was recorded at a temperature of 563~K and
the superlattice reflections are now absent. This is strong evidence
that they originate from antiferromagnetism. 

At room temperature after heating above $T_N$, the pattern of
antiferromagnetic domains would be expected to change and this is what
was observed. The reflections at ${\bf Q}=\pm{1 \over
  2}(11\overline{1})$ were no longer present in the region from which
Fig.~\ref{fig1} was taken but an adjacent region 250~nm away showed
${\bf Q}=\pm{1 \over 2}(11\overline{1})$ reflections which had not
been there previously. Between these two regions was a line of
contrast of the type identified in ref.~\cite{Remaut65} as a T-type
antiferromagnetic domain wall.

The same experiment was performed in a different region of the sample
exhibiting reflections close to $\pm{1 \over 2}(\overline{3}11)$
(Supp. Info. 6). Even though antiferromagnetism is expected to produce
reflections at this wavevector, these did not disappear above $T_N$
and as discussed in Supp. Info. 6 it is likely these originate from a
surface reconstruction.

To compare the intensities of the $\pm{1 \over 2}(11\overline{1})$
superlattice reflections with theory and further rule out a surface
reconstruction as their cause, a series of diffraction patterns was
then taken on the [112] zone axis at different thicknesses, each from
an 80~nm diameter region. The thickness was found by taking a
bright-field image under two-beam conditions using the 000 and
$2\overline{2}0$ reflections to produce an image showing thickness
fringes where the change in thickness from one fringe to the next is
the extinction distance, $\xi_{2\overline{2}0}$. This was calibrated
using the standard two-beam convergent-beam technique described in
ref.~\cite{Williams09} to give $\xi_{2\overline{2}0}=66 \pm 2$~nm.

\begin{figure}
\includegraphics[width=80mm]{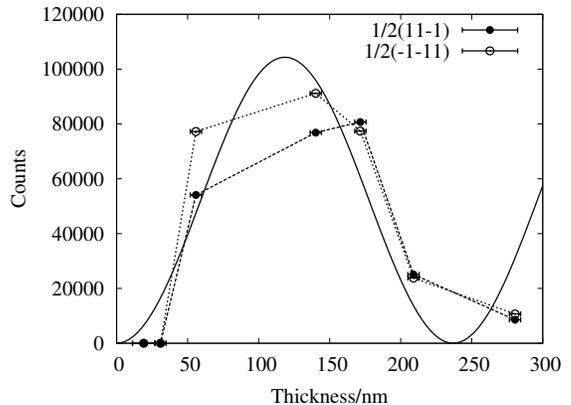}
\caption{\label{fig2} Intensity versus thickness for the $\pm{1 \over
    2}(11\overline{1})$ reflections. The solid line shows the expected
  oscillation in the intensity.}
\end{figure}

The data points in Fig.~\ref{fig2} show the variation of the intensity
of the $\pm{1 \over 2}(11\overline{1})$ reflections as a function of
thickness. The $x$-errorbars show the uncertainty due to the
extrapolation between thickness finges. The error in the measured
intensities is very small and the principal cause of data scatter is
specimen bending which changes the deviation parameter so that the
$\pm{1 \over 2}(11\overline{1})$ reflections do not have the same
intensity.  

The behaviour in Fig.~\ref{fig2} is not consistent with a surface
reconstruction. For a surface reconstruction, the intensity should be
greatest at the smallest thickness and diminish as the thickness
increases. Here the intensity is zero at zero thickness and oscillates
as the thickness increases as expected for a reflection originating
from the bulk.

The deviation parameter for the $\pm{1 \over 2}(11\overline{1})$
reflections on the [112] zone axis is $s=4.23 \times
10^{-4}$~\AA$^{-1}$ so the intensity should oscillate with thickness
with period $1/s=237$~nm. This prediction is shown by the solid line
in Fig.~\ref{fig2} and the only fitting parameter used was the
amplitude of the oscillation. It can be seen that the data points
follow this trend.

Eqn.~\ref{I} predicts the intensity ratio of the antiferromagnetic ${1
  \over 2}(11\overline{1})$ beam relative to the 000 beam but makes
use of the phase object approximation and so will only apply to the
thinnest parts of the specimen. There are several comparisons one
could make with experiment.

The simplest is to look at the diffraction pattern from the thinnest
region where superlattice reflections could be discerned, 57~nm. The
measured intensity ratio here is $1.25\times 10^{-4}$ for ${1 \over
  2}(11\overline{1})$ and $1.79 \times 10^{-4}$ for ${1 \over
  2}(\overline{11}1)$. The predicted ratio (accounting for the
deviation parameter) is $0.50 \times 10^{-4}$. Alternatively one could
state that the theory neglects the loss of intensity of the 000 beam
due to scattering to the structural reflections and therefore one
should divide not by the intensity of the 000 beam for 57~nm but by
the intensity of 000 for the thinnest region at which data was
recorded. This gives ratios of $0.54 \times 10^{-4}$ and $0.77 \times
10^{-4}$, closer to the predicted ratio of $0.50 \times
10^{-4}$. Finally one could use the fitted curve to give the intensity
ratio for the thinnest region of the sample from which data was
recorded, 19~nm. This gives a ratio of $6.4 \times 10^{-6}$ as
compared with a predicted ratio of $6.6 \times 10^{-6}$.

We consider the combined facts that the superlattice reflections are
at the correct positions, originate from the bulk, disappear above the
N\'{e}el temperature and have intensities in the range predicted by
theory compelling evidence that they originate from
antiferromagnetism.

\begin{figure}
\includegraphics[width=60mm]{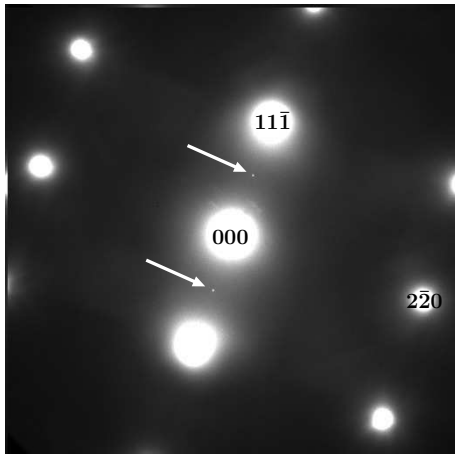}
\caption{\label{fig3} Electron diffraction pattern taken from the
  [112] zone axis in a field of less than 10~mT. Antiferromagnetic
  reflections (indicated by arrows) are seen at $\pm{1 \over
    2}(11\overline{1})$.}
\end{figure}

The above experiment was conducted with the microscope in its normal
operating condition with a field of 2.8~T applied to the specimen so
that its spins flop. Fig.~\ref{fig3} was taken from the same area
using a `Lorentz lens' instead of the objective lens so that the
sample was in a field less than 10~mT. The same superlattice
reflections are observed. Saito {\it et al.}~\cite{Saito80} showed
that if the field inducing the spins to flop is removed, they return
to point along their easy axes. Since the superlattice reflections are
still visible in Lorentz mode, they must be aligned along either
$[2\overline{1}1]$ or $[\overline{1}21]$ but not [112] (i.e. parallel
to the electron beam) as they would then be invisible. The intensity
ratios are $2.29 \times 10^{-4}$ and $2.74 \times 10^{-4}$ for the ${1
  \over 2}(11\overline{1})$ and ${1 \over 2}(\overline{11}1)$
reflections respectively and the diffraction pattern comes from a
region $119 \pm 10$~nm thick. Supp. Info. 3 predicts an intensity
ratio of $1.06 \times 10^{-4}$ in the flopped configuration and $0.79
\times 10^{-4}$ for the easy-axis alignment. The predicted intensities
for the flopped and easy-axis configurations are consistent with
experiment but too similar to confirm the orientation of the spins.

It would be advantageous to develop a technique for imaging
antiferromagnetism using electron microscopy. We have attempted dark
field imaging but the images were dominated by diffuse scattering
as the antiferromagnetic reflections are so weak. We have also
attempted high resolution imaging which would give information on an
atomic scale but with no success so far. We shall soon be conducting
simulations to ascertain the effects of antiferromagnetism on high
resolution images.

The most promising imaging technique is spatially resolved diffraction
where a narrow, near-parallel electron beam is rastered across the
specimen and a diffraction pattern recorded at each point. An image
can be built up by plotting the intensity of a reflection as a
function of position. This is not standard on most microscopes and its
implementation will require modifying the software on our
microscope. In our experiments, diffraction patterns showing
antiferromagnetic refelections were obtained from regions as small as
10.5~nm in diameter. Tao {\it et al.}~\cite{Tao09} have used this
technique to image charge-ordered domains in
La$_{0.55}$Ca$_{0.45}$MnO$_3$ with a resolution of 1.7~nm.

In summary, antiferromagnetic reflections can be observed in
transmission electron diffraction patterns from NiO with an intensity
$\sim 10^4$ times less intense than the structural reflections. We
give a model to predict the intensities of the antiferromagnetic
reflections which agrees with our observations, allowing deductions to
be made about the direction of the magnetic moments. The diffraction
patterns taken here came from regions as small as 10.5~nm in diameter
and such patterns could be used to image the antiferromagnetic
structure using spatially resolved electron diffraction with a
nanometre resolution.


\section{Supplementary Information 1: Estimate of the Force on an Electron as it Passes Through an Antiferromagnet}

For transmission electron microscopy, the Lorentz force is about 500
times larger than the force the electron feels due to its magnetic
dipole moment as the following estimate shows. The maximum Lorentz
force, ${\bf F_e}=-e{\bf v} \times {\bf B}$ felt by an electron
passing through an antiferromagnet is about $2.4\times 10^{-11}$~N
(using $v=2.3 \times 10^8$~ms$^{-1}$ for 300~keV electrons and
$B=\mu_0M_0=0.64$~T where $M_0$ is the sublattice magnetisation for
NiO) whereas the force due to the electron's magnetic moment is ${\bf
  F_\mu}={\bm \upmu}_e. \nabla {\bf B}$ and has a maximum value of
$F_\mu\approx\mu_B(2B/d)=4.9\times10^{-14}$~N (where $d=2.4$~\AA\ is
the spacing between the (111) planes of antiparallel spins in NiO). A
similar estimate shows that the energy of the electron beam would need
to be 0.7~eV before the Lorentz and dipolar forces are equal.

In fact, unlike neutron diffraction, the dipolar force never dominates
and at low energies, the dominant force is from the exchange
interaction arising from the fact that the electrons in the beam and
the antiferromagnet are identical particles as discussed by DeWames
and Vredevoe~\cite{DeWames67}. The exchange interaction is larger than
the dipole interaction by a factor of the rest mass energy of the
electron (511~keV) divided by the kinetic energy of an electron in the
beam~\cite{DeWames67}. This ratio is 1.7 at 300~keV and so it is the
Lorentz interaction which dominates at high energies.

Low energy (32~eV) electron diffraction patterns have been acquired
from the first atomic layer of an antiferromagnet by Menon {\it et
  al.}~\cite{Menon11} and here the exchange interaction dominates the
scattering process as it is about 16000 times greater than the dipolar
interaction and 2300 times greater than the Lorentz force. Menon {\it
  et al.}~\cite{Menon11} have used these diffraction patterns to image
antiferromagnetic domains in the first atomic layer of the sample with
a resolution of 10~nm.

\section{Supplementary Information 2: Antiferromagnetic Domains}

\begin{figure}[h]
\includegraphics[width=65mm]{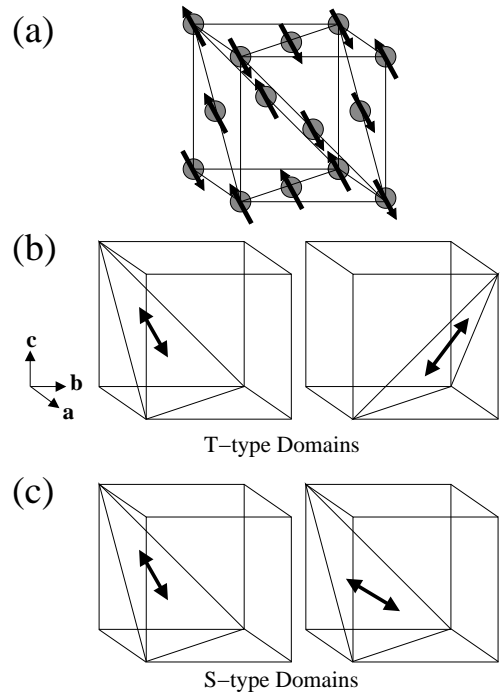}
\caption{\label{dom_wall_types} (a) Arrangement of Ni ions and spins
  in one unit cell of NiO showing the ferromagnetic alignment in the
  (111) planes. (b) Changes in the direction of the magnetic moments
  due to a T-type and (c) an S-type domain boundary. The double headed
  arrows indicate the direction of the magnetic moments in the
  ferromagnetic (111) planes shown.}
\end{figure}

In NiO, the magnetic moments are aligned ferromagnetically along one
set of $\{111\}$ planes with the moments in alternating planes being
antiparallel as shown in Fig.~\ref{dom_wall_types}(a). There are two
types of antiferromagnetic domain boundary: a twin (T-type) boundary
occurs when the magnetic order changes to a different set of (111)
planes and the accompanying distortion generates a crystallographic
twin (Fig.~\ref{dom_wall_types}(b)). The spin (S-type) domain boundary
occurs when the same set of (111) planes remains ferromagnetic but the
magnetic moments point in a different direction
(Fig.~\ref{dom_wall_types}(c)).

\section{Supplementary Information 3: Theory of Electron Diffraction from an Antiferromagnet}
\label{Theory}

Here we calculate the appearance of an electron diffraction pattern from an
antiferromagnet and estimate the intensities of the antiferromagnetic
reflections. To do this, we arrange Cartesian coordinates so that the electron
beam is travelling in the $z$-direction before it hits the specimen and then
the wavefunction of the electron $\psi(x,y,z)$ at any point can be found using
the Schr\"{o}dinger equation for fast electrons \cite{Kirkland98,Pozzi89}:

\begin{equation}
{\partial \psi \over \partial z}=i{\lambda \over 4\pi}\nabla_{xy}^2\psi+i\left(C_EV-{2\pi e \over h}A_z\right)\psi-{\lambda e \over h}{\bf A}.\nabla\psi
\label{eq:a}
\end{equation}

where $V(x,y,z)$ is the electrostatic potential, ${\bf A}(x,y,z)$ is
the magnetic vector potential, $\lambda$ the electron wavelength and
$C_E={2\pi e\over \lambda}\left({E+mc^2 \over E(E+2mc^2)}\right)$
where $E$ is the kinetic energy of the electrons, $m$ the rest mass of
the electrons and $c$ is the speed of light in a vacuum.

This equation is used throughout electron microscopy to simulate
images and diffraction patterns but usually with only the
electrostatic terms. Here we have retained the magnetic terms and
inserting a suitable expression for the vector potential from an
antiferromagnet would allow very accurate simulations of images and
diffraction patterns. This equation can only be solved iteratively,
however, usually using the multislice or Bloch wave
approaches~\cite{Kirkland98} and to give an analytic solution, we make
the phase object approximation where only the second term on the right
hand side equation \ref{eq:a} is retained. This approximation is valid
for very thin specimens but gives a physical insight into the
situation and provides an estimate of the intensities expected in a
diffraction pattern. Having made this approximation, equation
\ref{eq:a} is readily solved to give the wavefunction of the electron
as it emerges from a sample of thickness $t$ (called the `exit-plane
wavefunction') as $\psi(x,y)=e^{i\phi(x,y)}$ where

\begin{equation}
\phi(x,y)=\int_{-t/2}^{t/2}\left(C_EV(x,y,z)-{2\pi e \over h}A_z(x,y,z)\right)\mathrm{d}z
\end{equation}

(We have assumed that both potentials are zero outside the specimen
which is very reasonable for an antiferromagnet.)

We now show that the magnetic contribution to the phase is small by
taking the case where the antiferromagnetic wavevector ${\bf q}$
points in the $x$-direction and the flux density oscillates in
$y$. This will give the largest possible phase shift as the B-field is
normal to the electron beam direction $z$. If we write
$B_y=B_0\cos(2\pi q_xx)$, a suitable vector potential is
$A_z=-(B_0/2\pi q_x)\sin(2\pi q_xx)$ and this gives a phase shift
$\phi(x)=(eB_0t/hq_x)\sin(2\pi q_xx)$. For NiO, $B_0=0.64$~T and
$q=\sqrt(3)/2a=0.207$~\AA$^{-1}$, the maximum phase shift for a 100~nm
thick specimen is $7.5\times 10^{-3}$ radians.

Returning to the general case, this means that we can make the weak phase
object approximation for the magnetic contribution to the phase, $\exp[-i 2\pi
  e /h\int_{-\infty}^\infty A_zdz]\approx 1-i2\pi e/h \int_{-\infty}^\infty
A_zdz$ and if the electrostatic contribution to the wavefunction is denoted
$\psi_V\equiv\int_{-\infty}^{\infty}\exp[iC_EV(x,y,z)dz]$, the exit-plane
wavefunction is

\begin{eqnarray}
\psi(x,y)&=&\psi_V\left(1-i{2\pi e \over h}\int_{-t/2}^{t/2} A_zdz\right)\nonumber\\
&=&\psi_V\left(1-i{2\pi e \over h}\int_{-\infty}^{\infty} A_zh(z)dz\right)
\end{eqnarray}

where $h(z)$ is a top hat function which is 1 for $-t/2<z<t/2$ and
zero otherwise.

The diffraction pattern is the squared modulus of the Fourier transform of the
exit plane wavefunction, $|\Psi(k_x,k_y)|^2$, and so, using the Fourier
transform convention $\Psi({\bf k})=\int_{-\infty}^\infty \psi({\bf r})e^{2\pi
  i {\bf k.r}}\,\mathrm{d}^2{\bf r}$, we obtain

\begin{eqnarray}
\Psi(k_x,k_y)&=&\Psi_V*\bigg(\delta(k_x)\delta(k_y)-i{2\pi e \over h}\nonumber\\
&\times&\big(\widetilde{A}_z(k_x,k_y,k_z)*t{\rm sinc}(\pi k_zt)\big)_{k_z=0}\bigg)
\end{eqnarray}

where $\widetilde{A}_z$ denotes the three dimensional Fourier
transform of $A_z$, ${\rm sinc(X)}\equiv \sin(X)/X$, $*$ denotes a
convolution and we have made use of the convolution and Fourier
projection theorems.

We now calculate the magnetic vector potential for an antiferromagnet
basing our method on that used to derive the intensities for neutron
diffraction described in ref~\cite{Marshall71}. The result could be
used in equation~\ref{eq:a} but here we apply it for the phase object
approximation. If we assume that the contribution from the orbital
angular momentum of the electrons in the antiferromagnet is quenched,
the magnetic vector potential can be found by summing the fields due
to the magnetic moments generated by the spin of the unpaired
electrons on each atom. A single electron at position ${\bf R}_i$ with
magnetic moment ${\bm \upmu}_i$ will generate a vector potential at
position ${\bf r}$ of

\begin{equation}
{\bf A}_i({\bf r})={\mu_0 \over 4\pi}{{\bm \upmu}_i \times ({\bf r}-{\bf R}_i) \over |{\bf r}-{\bf R}_i|^3}={\mu_0 \over 4\pi}{{\bm \upmu}_i \times {\bf r} \over r^3}*\delta({\bf r}-{\bf R}_i)
\end{equation}

The vector potential for the entire crystal is then the sum over all
the unpaired electrons

\begin{equation}
{\bf A}({\bf r})=\sum_i{\mu_0 \over 4\pi}{{\bm \upmu}_i \times {\bf r} \over r^3}*\delta({\bf r}-{\bf R}_i)
\end{equation}

For a simple antiferromagnet where all the electron spins are
collinear, we can write ${\bm \upmu}_i=\widehat{\bm \upmu}\mu_i$ where
$\widehat{\bm \upmu}$ is a unit vector pointing in the direction of
the magnetic moment and $\mu_i$ expresses the magnitude and sign of
the moment. Then

\begin{equation}
{\bf A}({\bf r})={\mu_0 \over 4\pi}{\widehat{\bm \upmu} \times {\bf r} \over r^3}*\sum_i\mu_i\delta({\bf r}-{\bf R}_i)
\end{equation}

We recognise the last term as the local magnetisation, $M({\bf r})$ so
we can write:

\begin{equation}
{\bf A}({\bf r})={\mu_0 \over 4\pi}{\widehat{\bm \upmu} \times {\bf r} \over r^3}*M({\bf r})
\end{equation}

The magnetisation is the product of the electron number density and the
magnetic moment of each electron $M({\bf r})=\rho({\bf r})\mu({\bf
  r})$. To model an antiferromagnet, we allow the magnetisation to
vary depending on its position in the crystal. For NiO, this is
conventionally done by using a `magnetic unit cell' which is twice the
size of the structural unit cell in each of the $a$, $b$ and $c$
directions and reversing the sign of ${\bm \upmu}$ on alternating
(111) planes~\cite{Roth58}. Instead, we use the same unit cell as the
structural unit cell but allow the magnetisation to vary as $M({\bf
  r})=\rho_0({\bf r})\mu_B\cos(2\pi{\bf q.r})$ where $\rho_0({\bf r})$
is the electron number density in the absence of the modulation and
$\mu_B$ is the size of the magnetic moment on each electron, the Bohr
Magneton. This represents the lowest-order Fourier component of an
antiferromagnetic modulation and this approach has the advantages that
the wavevector of the modulation, ${\bf q}$, need not be commensurate
with the atomic lattice and that it can have any direction.

It turns out to be convenient to write the number density of electrons
in terms of the number density of electrons for one atom. If the atom is
labelled $j$ and the unpaired electron density with the origin at the
centre of the atom is denoted $\rho_j({\bf r})$, we can write:

\begin{eqnarray}
\rho_0({\bf r})&=&\sum_n\sum_j\rho_j({\bf r}-{\bf R}_n-{\bf R}_j)\nonumber\\
&=&\sum_n\sum_j\rho_j({\bf r})*\delta({\bf r}-{\bf R}_n-{\bf R}_j)
\end{eqnarray}

where ${\bf R}_j$ is the position of atom $j$ within its unit cell and
${\bf R}_n$ is the position of the unit cell within the crystal. Thus,
the vector potential becomes:

\begin{eqnarray}
{\bf A}({\bf r})&=&{\mu_0 \mu_B\over 4\pi}{\widehat{\bm \upmu} \times {\bf r} \over r^3}*\sum_n\sum_j(\rho_j({\bf r})*\delta({\bf r}-{\bf R}_n-{\bf R}_j))\nonumber\\
&\times&\cos(2\pi{\bf q.r})
\end{eqnarray}

The $z$-component of the vector potential can be selected by multiplying by a unit vector in the $z$-direction, $\widehat{\bf z}$ thus:

\begin{eqnarray}
A_z({\bf r})&=&{\mu_0\mu_B \over 4\pi}{\widehat{\bm \upmu} \times {\bf r} \over r^3}{\bf .\widehat z} *\sum_n\sum_j(\rho_j({\bf r})*\delta({\bf r}-{\bf R}_n-{\bf R}_j))\nonumber\\
&\times&\cos(2\pi{\bf q.r})
\end{eqnarray}

The Fourier transform of this expression is

\begin{eqnarray}
\widetilde{A}_z({\bf k})&=&{\mu_0 \over 4\pi k}(\widehat{\bm \upmu} \times \widehat{\bf k}){\bf .\widehat z}\sum_n\sum_j(\widetilde\rho_j({\bf k})e^{2\pi i{\bf k.R}_n}e^{2\pi i{\bf k.R}_j})\nonumber\\
&*&{\mu \over 2}(\delta({\bf k}-{\bf q})+\delta({\bf k}+{\bf q}))
\end{eqnarray}

where we have used the result that the Fourier transform of
$\left[{{\bf r}\over r^3}\right]$ is $-2\left({{\bf k}\over
  k^2}\right)$ and introduced a unit vector so that ${\bf
  k}=k\widehat{\bf k}$. Recognising that the Fourier transform of the
density of unpaired electrons is $n_j$, the number of unpaired
electrons associated with atom $j$ multiplied by the magnetic form
factor, $f_j(k)$ (identical to that used in neutron diffraction) we
can now write

\begin{eqnarray}
\widetilde{A}_z({\bf k})&=&-{\mu_0\mu_B \over 4\pi k}({\widehat{\bm \upmu} \times \widehat{\bf k}}){\bf .\widehat z}\sum_jn_jf_j({\bf k})e^{2\pi i{\bf k.R}_j}\sum_ne^{2\pi i{\bf k.R}_n}\nonumber\\
&*&(\delta({\bf k}-{\bf q})+\delta({\bf k}+{\bf q}))
\end{eqnarray}

Using the result that $\sum_ne^{2\pi i {\bf k.R}_n}={1 \over
  \Omega}\sum_{\bf G}\delta({\bf k}-{\bf G})$ where ${\bf G}$ is a
reciprocal lattice vector of the structural unit cell and $\Omega$ is
the unit cell volume, we obtain

\begin{eqnarray}
\widetilde{A}_z({\bf k})&=&-{\mu_0\mu_B \over 4\pi k}({\widehat{\bm \upmu} \times \widehat{\bf k}}){\bf .\widehat z}\sum_jn_jf_j({\bf k})e^{2\pi i{\bf k.R}_j}\nonumber\\
&\times&{1 \over \Omega}\sum_{\bf Q}\delta({\bf k}-{\bf Q})
\end{eqnarray}

where ${\bf Q}={\bf G} \pm {\bf q}$. So the Fourier transform of the exit-plane wavefunction is

\begin{eqnarray}
\Psi(&k_x&,k_y)=\Psi_V(k_x,k_y)*\Bigg(\delta(k_x)\delta(k_y)\nonumber\\
&&+i{e \over 2h}{\mu_0\mu_B \over \Omega}{t \over k}({\bm \widehat{\upmu}}\times{\bf \widehat{k}}).{\bf \widehat{z}}\sum_jn_jf_j({\bf k})e^{2\pi i{\bf k.R_j}}\nonumber\\
&&\times\sum_{\bf Q}\delta(k_x-Q_x)\delta(k_y-Q_y)\mathrm{sinc}(\pi Q_zt)\Bigg)
\end{eqnarray}

Thus if the structural Bragg reflections occur at ${\bf G}$, the
antiferromagnetic reflections occur at ${\bf Q}={\bf G}\pm {\bf q}$. It can be
seen that the scattering amplitude is proportional to a geometric factor
$G_e({\bf k})=((\bm {\hat{\upmu}}\times{\bf \hat{k}}).{\bf \hat{z}}/k)$
multiplied by the structure factor for magnetic scattering, $F({\bf
  k})\equiv\sum_jn_jf_j({\bf k})e^{2\pi i{\bf k.R_j}}$. For neutron
diffraction, there is a similar expression where the structure factor is the
same but the geometric factor $G_n({\bf k})\equiv\ |{\bf
  \widehat{k}}\times\bm{\upmu}\times {\bf \widehat{k}}|$. 

We use this expression for $\Psi(k_x,k_y)$ to calculate the
intensities of the antiferromagnetic reflections $I_{\bf Q}$ in an
electron diffraction pattern for a 100~nm thick specimen relative to
the 000 beam, $I_0$. The results are shown in table~\ref{table1} for
various directions of $\bm{\upmu}$. In its construction we made the
further approximation that the 000 beam is much stronger than the
other structural Bragg peaks (in the diffraction patterns acquired
here it was typically 3--4 times more intense than the neighbouring
peaks) so that as far as magnetic scattering is concerned, $\Psi_V$ is
simply a delta function at 000 and the intensity ratios are given by

\begin{equation}
{I_{{\bf Q}} \over I_0}=\left({e \over 2h}{\mu_0\mu_B \over \Omega}{t \over
    Q}\,(\bm {\widehat{\upmu}}\times {\bf \widehat{Q}}).{\bf \widehat{z}}\,F({\bf Q})\right)^2
\end{equation}

\section{Supplementary Information 4: Table of Calculated Intensities of Antiferromagnetic Reflections}

Table~\ref{table1} gives the intensities of the antiferromagnetic
reflections for 100~nm thick NiO for the $[112]$-type zone axes
investigated in this experiment calculated according to the method in
Supp. Info. 3. The form factors used in its construction were
$f\left({1 \over 2}(111)\right)=0.92 \pm 0.03$, $f\left({1 \over
  2}(113)\right)=0.82 \pm 0.02$ and $f\left({1 \over
  2}(333)\right)=0.58 \pm 0.02$ derived using neutron diffraction in
ref~\cite{Alperin61}. The symbols used are as follows: ${\bf z}$ is
the direction of the electron beam relative to the crystal, ${\bf q}$
is the wavevector of the antiferromagnetic modulation, ${\bf Q}$ gives
the reciprocal-space coordinates of the reflection being examined,
$\bm{\upmu}$ gives the direction of the magnetic moments and $I_{\bf
  Q}/I_0$ gives the intensity of the reflection relative to the
central beam. The intensity is given both for the case that
$\bm{\upmu}$ points in one of the [112]-type easy directions and for
the $[1\overline{1}0]$-type `flopped' directions.

\begin{table}[!]
\caption{\label{table1} Calculated Intensity Ratios for
  Antiferromagnetic Reflections in an Electron Diffraction Pattern
  from 100~nm thick NiO.}
\begin{ruledtabular}
\begin{tabular}{ccccc}
${\bf z}$&
${\bf q}$&
${\bf Q}$&
${{\bm \upmu}}$&
$I_{\bf Q}/I_0$\\[3pt]
\colrule\rule{0pt}{12pt}
$[211]$ & ${1 \over 2}(\overline{1}11)$ & ${1 \over 2}(\overline{1}11)$ & $[211]$ & 0\\
 & & & $[12\overline{1}]$ & $(1.40 \pm 0.08) \times 10^{-4}$\\
 & & & $[1\overline{1}2]$ & $(1.40 \pm 0.08) \times 10^{-4}$\\
 & & & $[01\overline{1}]$ & $(1.88 \pm 0.12) \times 10^{-4}$\\[3pt]
 & & ${1 \over 2}(\overline{3}33)$ & $[211]$ & 0\\
 & & & $[1\overline{2}1]$ & $(6.0 \pm 0.4) \times 10^{-6}$\\
 & & & $[1\overline{1}2]$ & $(6.0 \pm 0.4) \times 10^{-6}$\\
 & & & $[01\overline{1}]$ & $(8.4 \pm 0.4) \times 10^{-6}$\\[3pt]
 & ${1 \over 2}(111)$ & ${1 \over 2}(11\overline{3})$ & $[\overline{2}11]$ & $(2.64 \pm 0.12)\times 10^{-5}$\\
 & & & $[1\overline{2}1]$ & $(2.96 \pm 0.12)\times 10^{-5}$\\
 & & & $[11\overline{2}]$ & $(1.04 \pm 0.04)\times 10^{-7}$\\
 & & & $[01\overline{1}]$ & $(1.12 \pm 0.04)\times 10^{-5}$\\[3pt]
 & & ${1 \over 2}(1\overline{3}1)$ & $[\overline{2}11]$ & $(2.64 \pm 0.12)\times 10^{-5}$\\
 & & & $[1\overline{2}1]$ & $(1.04 \pm 0.04)\times 10^{-7}$\\
 & & & $[11\overline{2}]$ & $(2.96 \pm 0.12)\times 10^{-5}$\\
 & & & $[01\overline{1}]$ & $(1.12 \pm 0.04)\times 10^{-5}$\\[3pt]
\colrule\rule{0pt}{12pt}
$[121]$ & ${1 \over 2}(1\overline{1}1)$ & ${1 \over 2}(1\overline{1}1)$ & $[21\overline{1}]$ & $(1.40 \pm 0.08) \times 10^{-4}$\\
 & & & $[121]$ & 0 \\
 & & & $[\overline{1}12]$ & $(1.40 \pm 0.08) \times 10^{-4}$ \\
 & & & $[10\overline{1}]$ & $(1.88 \pm 0.12) \times 10^{-4}$ \\[3pt]
 & & ${1 \over 2}(3\overline{3}3)$ & $[21\overline{1}]$ & $(6.0 \pm 0.4) \times 10^{-6}$\\
 & & & $[121]$ & 0 \\
 & & & $[\overline{1}12]$ & $(6.0 \pm 0.4) \times 10^{-6}$ \\
 & & & $[10\overline{1}]$ & $(8.4 \pm 0.4) \times 10^{-6}$ \\[3pt]
 & ${1 \over 2}(111)$ & ${1 \over 2}(11\overline{3})$ & $[\overline{2}11]$ & $(2.96 \pm 0.16) \times 10^{-5}$\\
 & & & $[1\overline{2}1]$ & $(2.64 \pm 0.12) \times 10^{-5}$ \\
 & & & $[11\overline{2}]$ & $(1.04 \pm 0.04) \times 10^{-7}$ \\
 & & & $[10\overline{1}]$ & $(1.12 \pm 0.04) \times 10^{-5}$ \\[3pt]
 & & ${1 \over 2}(\overline{3}11)$ & $[\overline{2}11]$ & $(1.04 \pm 0.4) \times 10^{-7}$\\
 & & & $[1\overline{2}1]$ & $(2.64 \pm 0.12) \times 10^{-5}$ \\
 & & & $[11\overline{2}]$ & $(2.96 \pm 0.16) \times 10^{-5}$\\
 & & & $[10\overline{1}]$ & $(1.12 \pm 0.04) \times 10^{-5}$ \\[3pt]
\colrule\rule{0pt}{12pt}
$[112]$ & ${1 \over 2}(11\overline{1})$ & ${1 \over 2}(11\overline{1})$ & $[2\overline{1}1]$ & $(1.40 \pm 0.08) \times 10^{-4}$\\
 & & & $[\overline{1}21]$ & $(1.40 \pm 0.08) \times 10^{-4}$ \\
 & & & $[112]$ & 0 \\
 & & & $[1\overline{1}0]$ & $(1.88 \pm 0.12) \times 10^{-4}$ \\[3pt]
 & & ${1 \over 2}(33\overline{3})$ & $[2\overline{1}1]$ & $(6.0 \pm 0.4) \times 10^{-6}$ \\
 & & & $[\overline{1}21]$ & $(6.0 \pm 0.4) \times 10^{-6}$ \\
 & & & $[112]$ & 0 \\
 & & & $[1\overline{1}0]$ & $(8.4 \pm 0.4) \times 10^{-6}$ \\[3pt]
 & ${1 \over 2}(111)$ & ${1 \over 2}(\overline{3}11)$ & $[\overline{2}11]$ & $(1.04 \pm 0.04) \times 10^{-7}$\\
 & & & $[1\overline{2}1]$ & $(2.96 \pm 0.16) \times 10^{-5}$\\
 & & & $[11\overline{2}]$ & $(2.64 \pm 0.12) \times 10^{-5}$ \\
 & & & $[1\overline{1}0]$ & $(1.12 \pm 0.04) \times 10^{-5}$ \\[3 pt]
 & & ${1 \over 2}(1\overline{3}1)$ & $[\overline{2}11]$ & $(2.96 \pm 0.16) \times 10^{-5}$\\
 & & & $[1\overline{2}1]$ & $(1.04 \pm 0.04) \times 10^{-7}$\\
 & & & $[11\overline{2}]$ & $(2.64 \pm 0.12) \times 10^{-5}$ \\
 & & & $[1\overline{1}0]$ & $(1.12 \pm 0.04) \times 10^{-5}$ \\[3 pt]
\end{tabular}
\end{ruledtabular}
\end{table}

\newpage

\section{Supplementary Information 5: Optical Microscopy}

\begin{figure}[h]
\includegraphics[width=80mm]{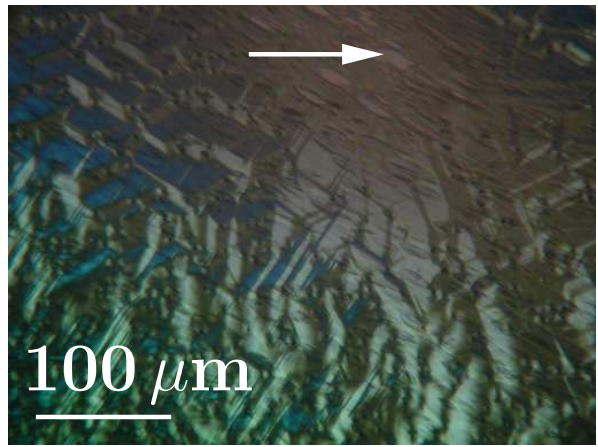}
\caption{\label{optical} Transmission optical micrograph taken with
  crossed polars looking down the [111] zone axis. The arrow shows a
  hole caused by ion-beam irradiation around which was the electron
  transparent material used for transmission electron microscopy.}
\end{figure}

We assessed the size of the antiferromagnetic domains using
optical microscopy. Fig.~\ref{optical} is a transmission optical micrograph
taken with crossed polars. This does not pick out the
antiferromagnetic domains {\it per se} but the twins associated with
each antiferromagnetic domain so only the T-type domains are
visible. (Kondoh {\it et al.}~\cite{Kondoh64} have reported that S-domains can
also be imaged with optical microscopy due to their associated lattice strain
but the contrast is much fainter.) This shows that the sizes of the T-domains
range from 2--80~$\upmu$m.

\section{Supplementary Information 6: Superlattice Reflections due to Surface Reconstruction}

We have shown that antiferromagnetic reflections can be seen in
electron diffraction patterns when the wavevector of the modulation
${\bf q}={1 \over 2}(11\overline{1})$ is normal to the electron beam,
producing reflections at ${\bf Q}=\pm{1 \over
  2}(11\overline{1})$. However, when ${\bf q}={1 \over 2}(111)$,
i.e. parallel to the thin direction of the crystal, superlattice
reflections should be produced at the $\pm{1 \over
  2}\{11\overline{3}\}$ positions.  We have acquired diffraction
patterns showing such superlattice reflections but have found that
they can also be generated by surface reconstructions which inevitably
occur on the (111) surfaces which terminate the sample.

\begin{figure}
\includegraphics[width=65mm]{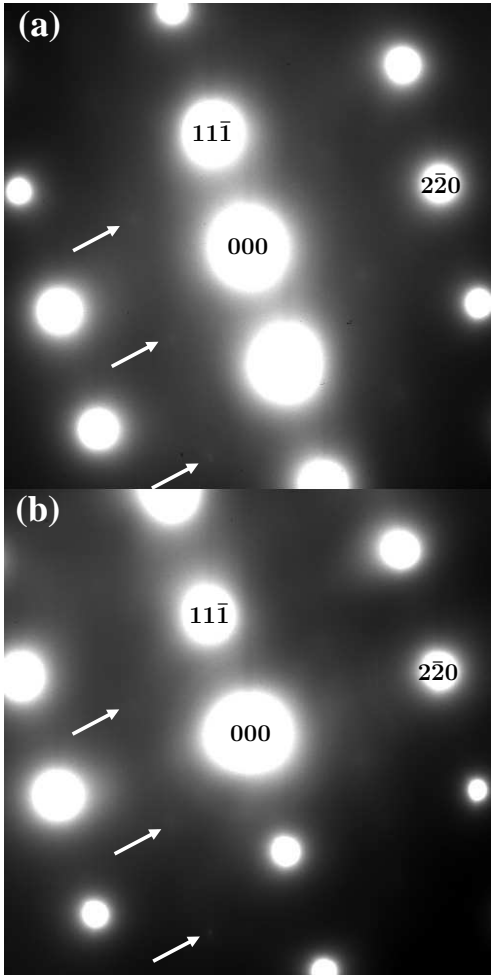}
\caption{\label{hotcold} (a) Electron diffraction pattern from the
  [112] zone axis taken at room temperature from a region of diameter
  880~nm with an exposure of 70~s showing superlattice reflections
  near $\pm{1 \over 2}(\overline{3}11)$, $\pm{1 \over
    2}(1\overline{3}1)$, $\pm{1 \over 2}(15\overline{3})$, $\pm{1
    \over 2}(51\overline{3})$ etc., some of which are indicated by
  arrows. (b) Electron diffraction pattern from the same region taken
  at 563~K, above the N\'{e}el temperature of 523~K, showing that the
  superlattice reflections are still present and are therefore not
  caused by antiferromagnetism.}
\end{figure}

Fig.~\ref{hotcold} shows the results of a heating experiment identical
to the that shown in Fig.~\ref{fig1} but from a different region of
the sample where ${1 \over 2}(\overline{3}11)$-type superlattice
reflections were observed. (a) was taken at room temperature and (b) at
563~K, above the N\'{e}el temperature of 523~K. In this case, however,
the superlattice reflections do not disappear in the high temperature
image showing that they are not related to antiferromagnetism.

\begin{figure}
\includegraphics[width=65mm]{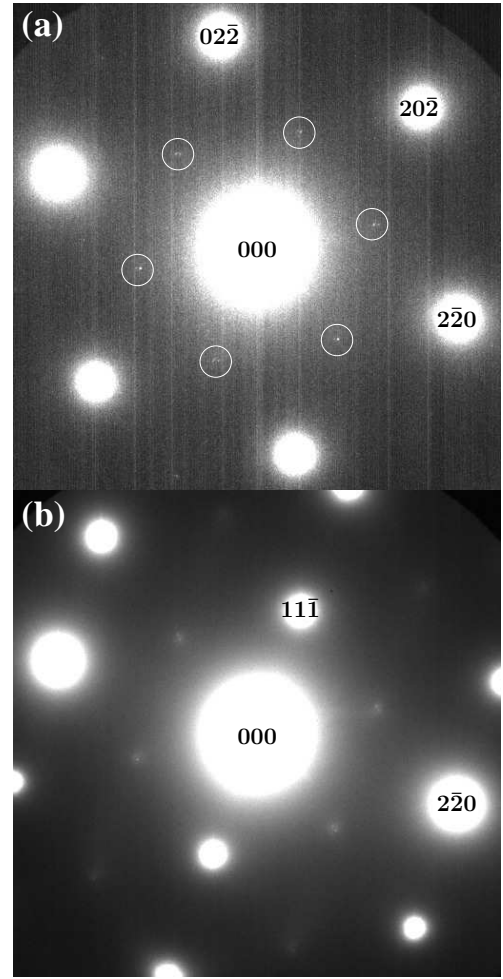}
\caption{\label{wrongzone} (a) Electron diffraction pattern from the
  [111] zone axis taken at room temperature with a 5~s
  exposure. Superlattice reflections (circled) can be seen near ${1
    \over 3}\{4\overline{22}\}$. (b) Electron diffraction pattern from
  the [112] zone axis taken from the same region as (a) with a 70~s
  exposure. Superlattice reflections can now be seen at ${1 \over
    2}\{\overline{3}11\}$.}
\end{figure}

Fig.~\ref{wrongzone} shows that the ${1 \over 2}(\overline{3}11)$-type
superlattice reflections are likely to be the result of a surface
reconstruction. (b) shows superlattice reflections near the ${1 \over
  2}(\overline{3}11)$-type positions in the [112] zone axis but (a)
shows very similar reflections in the [111] zone axis near the ${1
  \over 3}\{4\overline{22}\}$ positions. A surface reconstruction is
only about 2 monolayers thick and so will produce a relrod in
reciprocal space running in the (111) direction. This rod can cause
both the reflections in the [111] and [112] zone axes as ${1 \over
  3}(4\overline{22})+{1 \over 6}(111)={1 \over 2}(3\overline{11})$.

Ciston {\it et al.}~\cite{Ciston10} have investigated the surface
ordering in NiO using transmission electron diffraction using an
ultra-high vacuum microscope which allows {\it in situ}
annealing. Prior to annealing, their samples were thinned in the [111]
direction using a very similar method to that used here. The authors
do not report whether any surface order was observed prior to
annealing but show that after annealing at temperatures between 1220
and 1470~K, two types of reconstruction were produced. Both the
$\sqrt{3}\times\sqrt{3}$R$30^\circ$ and p$2\times 2$ reconstructions
produce reflections at ${1 \over 3}\{4\overline{22}\}$-type positions
in the [111] zone but they also produce additional reflections not
observed in the present experiments. In addition the reflections
observed here are frequently split, indicating an incommensurate
surface reconstruction whereas those observed by Ciston {\it et
  al.}~\cite{Ciston10} are all commensurate. The splitting of the
${\bf Q}={1 \over 2}(11\overline{3})$-type reflections seen in the
[112]-type zone axes, $\Delta{\bf Q}$, ranged in magnitude from
0--0.07$a^*$ (where $a^*$ is the reciprocal lattice vector) and there
seems to be no restriction on the angle of $\Delta{\bf Q}$ with
respect to the rest of the pattern.

The characteristics of the superlattice reflections caused by surface
ordering are very different to those of the ${1 \over
  2}(11\overline{1})$-type antiferromagnetic reflections described in
the main paper. The antiferromagnetic reflections are very sharp
whereas those from the surface reconstruction are frequently
streaked. Also, the intensity of the antiferromagnetic reflections
falls very rapidly with increaing wavevector so that, for example, ${1
  \over 2}(33\overline{3})$ is very much dimmer than ${1 \over
  2}(11\overline{1})$ and the same should be true for the ${1 \over
  2}(\overline{3}11)$-type reflections. In contrast, the higher order
reflections which can be seen in Figs.~\ref{hotcold} and
\ref{wrongzone} resulting from the surface reconstruction have a
similar intensity to the low order reflections.

Whilst we believe that some of the diffraction patterns we have
recorded do show antiferromagnetic reflections at the ${1 \over
  2}(\overline{3}11)$-type positions, we are not able to demonstrate
this definitively. In the future, we shall look at samples terminated
at a stable surface, preferably (110) although this has the slight
disadvantage that only two of the three possible ${1 \over
  2}(111)$-type superlattice reflections will be accessible. None of
this alters the main conclusion that the effects of antiferromagnetism
can be observed in electron diffraction patterns.

\begin{acknowledgments}
This work was funded by the Royal Society.
\end{acknowledgments}

\bibliography{NiO_2nd_revision}

\end{document}